\documentclass[aps]{revtex4}
\usepackage{amsfonts}
\usepackage{graphics}
\usepackage{graphicx}
\newcommand{\be}{\begin{equation}}
\newcommand{\ee}{\end{equation}}
\newcommand{\ben}{\begin{eqnarray}}
\newcommand{\een}{\end{eqnarray}}
\usepackage{amsmath}
\usepackage[latin1]{inputenc}
\begin{document}
\title{3D black holes on a 2-brane in 4D Minkowski space}
\author{D. Bazeia$^{a}$, F. A. Brito$^{b}$ and F. G. Costa$^{a,c}$}
\email{bazeia@fisica.ufpb.br, fabrito@df.ufcg.edu.br, geraldo.costa@ifrn.edu.br}
\affiliation{$^{a}$ Departamento de F\'{\i}sica, Universidade Federal da Para\'\i ba, Caixa Postal 5008, 58051-970 Jo\~ao Pessoa, Para\'{\i}ba, Brazil \\
$^{b}$ Departamento de F\'{\i}sica, Universidade Federal de Campina Grande, Caixa Postal 10071, 58109-970 Campina Grande, Para\'{\i}ba, Brazil\\
$^{c}$ Instituto Federal do Rio Grande do Norte-IFRN, 59380-000 Currais Novos, Rio Grande do Norte, Brazil}
\begin{abstract}

We investigate three-dimensional black hole solutions in the realm of pure and new massive gravity in 2+1 dimensions induced on a 2-brane embedded in a flat four-dimensional spacetime. There is no cosmological constant neither on the brane nor on the four-dimensional bulk. Only gravitational fields are turned on and we indeed find vacuum solutions as black holes in 2+1 dimensions even in the absence of any cosmological solution. There is a crossover scale that controls how far  the
three- or four-dimensional gravity manifests on the 2-brane. Our solutions also indicate that local BTZ and $SdS_3$ solutions can flow to local four-dimensional Schwarzschild like black holes, as one probes from small to large distances, which is clearly a higher dimensional manifestation on the 2-brane. This is similar to the DGP scenario where the effects of extra dimensions for large probed distances along the brane manifest.

\end{abstract}

\maketitle
\pretolerance10000

\section{Introduction}

The interest on 2+1 dimensional gravity has been renewed recently \cite{bht}. This in part is due to the difficulty to understand the quantization of four dimensional gravity, and lower dimensional gravity provides good scenarios for investigating important issues in the sense of shedding new light on higher dimensional gravity. Furthermore, holographic aspects of three-dimensional gravity has also been considered in the $AdS_3/CFT_2$ correspondence \cite{st1,st2}. On the other hand, pure 2+1 gravity is a topological theory and has no propagating gravitons. However, massive propagating gravitons can appear in some proposals, among them, the case where one adds gravitational Chern-Simons terms as first considered in \cite{CS}, which has been named topologically massive gravity (TMG), and also within the more recent context, where new massive gravity (NMG) has been put forward by Bergshoeff, Hohm and Townsend \cite{bht} --- see also \cite{denis-guara}. One may refer to these cases as `massive gravity', though here we shall focus only on the NMG case.

In the current study we search for new three-dimensional black hole solutions. The three-dimensional black holes in 2+1 dimensional gravity with a cosmological constant such as BTZ black holes have been first found long ago \cite{mann,banados,btz}. They are equivalent to $AdS_3$ spaces because they requires a negative cosmological constant. However, in the NMG context, one has also recently found a class of interesting solutions --- see e.g.  \cite{Ahmedov:2011yd}.

We focus on elaborating a connection between three- and four-dimensional gravity, if they can somehow flow to one another.
This may be related to the recent Ho\v rava program \cite{h1,h2} of considering renormalizable gravity as a theory that can flow effectively from four- to two-dimensional spacetime. Thus, in a similar direction we shall follow the alternate proposal of applying a DGP-like scenario \cite{DGP}. One of the first consequences of this procedure is the fact that now there exist black hole solutions even in the  pure three-dimensional gravity induced on a 2-brane embedded into a four-dimensional Minkowski space. For the new massive gravity, new black role solutions with distinct horizons are also found. Another important observation is that at large distance, in the IR regime, the three-dimensional solution approaches a four-dimensional Schwarzschild solution, whereas on the the other side, for small distances, i.e., in the UV regime, BTZ and three-dimensional de Sitter - Schwarzschild ($SdS_3$) black holes appear, the three-dimensional gravity is recovered and then an improved UV regime is achieved. One should also mention that there are well-known exact solutions describing black holes on a 2-brane embedded in four-dimensions found for branes in the Randall-Sundrum-like scenario \cite{rs,ehm}. 

In the current work we consider an alternative study, in which we analyze a 2+1-dimensional theory
of gravity as a 2-brane with induced gravity through a scalar curvature term. This is
the four-dimensional counterpart of the DGP scenario, which was originally developed in five-dimensions \cite{DGP}.
We take advantage of this set up to include the massive gravity term as it appears in the NMG context.

\section {Circularly Symmetric Black Hole Solution in 3D induced Gravity on a 2-brane in 4D flat spacetime}

In this section we consider a four-dimensional set up which is analogous to that employed in the DGP scenario in five-dimensions. In the current investigation, the theory is constructed based on a 2-brane with a scalar curvature term embedded in a flat
four-dimensional spacetime. The full action can be separated into two parts
\be S=S_{(4)}+S_{(3)},\ee
where
\be S_{(4)}=-\frac{1}{2\kappa_{4}^{2}}\int
d^{4}x\sqrt{|g|}\left(R^{(4)}-2\kappa_{4}^{2}L_{m}^{(4)}\right),
\ee
with $\kappa_{4}$ standing for the four-dimensional gravitational coupling, $ds^2_4=g_{ab}(r,z)dx^adx^b\ (a,b=0,1,2,3)$ being the metric of the four-dimensional spacetime 
and
\be S_{(3)}=-\frac{1}{2\kappa_{3}^{2}}\int
d^{3}x\sqrt{|q|}\left(R^{(3)}-2\kappa_{3}^{2}L_{m}^{(3)}\right),
\ee
with $\kappa_{3}$ standing for three-dimensional gravitational coupling and $ds^2_3=q_{\mu\nu}(r)dx^\mu dx^\nu\ (\mu,\nu=0,1,2)$ being the induced metric on the 2-brane at
$z=0$ with $q_{\mu\nu}(r)\equiv g_{\mu\nu}(r,z\!=\!0)$. It follows that the induced scalar curvature $R^{(3)}$ is made out of this three-dimensional metric and has no dependence with the transversal fourth coordinate $z$. The Einstein equations in four-dimensions for this theory are
\be G_{ab}=\kappa_{(4)}^{2}S_{ab},\ee
where
\be S_{ab}=T_{ab}+U_{ab}.\ee
$T_{ab}$ is the energy-momentum tensor for the matter fields and $U_{ab}$
is given by the curvature due to the induced metric on the 2-brane. The tensor
$T_{ab}$ can be written as
\be T_{ab}=T_{ab}|_{bulk}+T_{ab}|_{brane}.\ee

Once we shall only investigate vacuum solutions for this theory, the matter field Lagrangians $L_{m}^{(3)}=0$ and $L_{m}^{(4)}=0$, so that we have
\be T_{ab}=0.\ee
As a consequence, the Einstein equations become
\be G_{ab}=\kappa_{(4)}^{2}U_{ab},\ee 
where
\be U_{a}^{b}=\delta(z){\rm diag\ }(-\rho_{curv},p_{curv},p_{curv},0)\ee defines the energy and pressure of the brane curvature. 
It has been considered as the component of a `cosmic fluid' \cite{defayet2000}. In searching for three-dimensional black hole solutions embedded in a four-dimensional spacetime we shall use the following Ansatz
\be\label{4d-metric}
ds^{2}_{4}=-A(r,z)dt^2+\frac{1}{A(r,z)}dr^2+r^2d\theta^2+dz^2.\ee

In order to compute the components of the Einstein tensor, we take dot to represent derivative with respect to $r$, whereas prime stands for derivative with respect to $z$. Thus, the components are given by
\be
G_{t}^{t}=\frac{3}{4}\left(\frac{A'}{A}\right)^{2}+\frac{1}{2r}\dot{A}-\frac{1}{2}\frac{A''}{A},\ee
\be
G_{r}^{r}=-\frac{1}{4}\left(\frac{A'}{A}\right)^{2}+\frac{1}{2r}\dot{A}+\frac{1}{2}\frac{A''}{A},\ee
\be
G_{\theta}^{\theta}=\frac{1}{4}\left(\frac{A'}{A}\right)^{2}+\frac{1}{2}\ddot{A},\ee
\be
G_{z}^{z}=-\frac{1}{4}\left(\frac{A'}{A}\right)^{2}+\frac{1}{2}\ddot{A}+\frac{1}{r}\dot{A}.\ee

In the absence of the 2-brane the Einstein equations satisfy the bulk vacuum $G_{ab}=0$. It follows that using
\be G^{t}_{t}+G^{r}_{r}=0,\ee
we find the useful bulk equation for $A(r,z)$
\be \label{eq.4d}
\frac{1}{2}\left(\frac{A'}{A}\right)^{2}+\frac{1}{r}\dot{A}=0.\ee
We shall reintroduce the brane shortly through boundary conditions at $z=0$, such that ${A}(r,z)|_{z=0}\equiv{A}_0(r)$ and $\dot{A}(r,z)|_{z=0}\equiv\dot{A}_0(r)$. See discussion below.

Let us now obtain the junction condition across the thin 2-brane located at $z=0$. The
extrinsic curvature is given by
\be K_{\mu\nu}=q_{\mu}^{\alpha}\nabla_{\alpha}n_{\nu},\ee
where $n^{\alpha}=(0,0,0,1)$. For the metric we have previously assumed we have that $K_{\mu\nu}$ is
given by
\be \label{K}
K_{b}^{a}=\left(\frac{A'}{2A},-\frac{A'}{2A},0,0\right).\ee

The relation between $K_{ab}$ and the energy-momentum tensor is given by the Israel's junction condition
in the first derivative across the brane at $z=0$ $[A']=A'(0^+)-A'(0^-)$. If $A'(0^-)=-A'(0^+)$ such that $[A']=2A'(0^+)$ we find \cite{defayet2000}
\be
[K_{ab}]=K_{ab}(0^+)=-\frac{\kappa^{2}_{(4)}}{2}\left(U_{ab}-\frac{1}{2}Uq_{ab}\right).\ee
Thus using (\ref{K}) we find
\be
K_{r}^{r}(0^+)=-\frac{A'(0^+)}{2A_0}=-\frac{\kappa^{2}_{(4)}}{2}\left(p_{curv}-\frac{1}{2}(\rho_{curv}+2p_{curv})\right),
\ee
which implies that
\be \label{Kcurv}
\frac{A'(0^+)}{A_0}=-\frac{\kappa^{2}_{(4)}}{2}\rho_{curv}.
\ee

On the 2-brane the Einstein equations govern the induced metric $q_{\mu\nu}$ such that we can identify the induced energy-momentum tensor for the curvature brane as $U_{ab}=\delta(z)U_{\mu\nu}$. Recall that $U_{\mu\nu}$ has no dependence on the coordinate $z$. Now we can write the Einstein equations for the induced metric as follows
\be 
G_{\mu\nu}=-\kappa_{(3)}^{2}U_{\mu\nu}, \qquad \mu,\nu=0,1,2.
\ee
This allows us to obtain the following components
\be
U_{t}^{t}=U_{r}^{r}=-\frac{1}{\kappa_{(3)}^{2}}\frac{\dot{A}_0}{2r},\ee
and
\be
U_{\theta}^{\theta}=-\frac{1}{\kappa_{(3)}^{2}}\frac{\ddot{A}_0}{2}.\ee
Finally, we are able to recognize the energy density and the pressure for the brane curvature as follows
\be \label{curv}
\rho_{curv}=\frac{1}{\kappa_{(3)}^{2}}\frac{\dot{A}_0}{2r},\ee
and
\be p_{curv}=-\frac{1}{\kappa_{(3)}^{2}}\frac{\dot{A}_0}{2r}.\ee
Notice that $p_{curv}=-\rho_{curv}$, which shows that the `cosmic fluid' acts as a cosmological constant. This
signalizes the possibility of finding a three-dimensional black hole on the brane even if we start with a theory
without any cosmological constant or matter fields. 

Now using Eqs.~(\ref{Kcurv}) and (\ref{curv}) we find
\be
\frac{A'(0^+)}{A_0}=-\frac{\kappa^{2}_{(4)}}{4\kappa_{(3)}^{2}}\frac{\dot{A}_0}{r}.\ee
Applying this solution into (\ref{eq.4d}) with the boundary condition at $z=0$ we find
\be \frac{1}{2}\left[
\frac{\kappa^{2}_{(4)}}{4\kappa_{(3)}^{2}}\frac{\dot{A}_0}{r}
\right]^{2}+\frac{\dot{A}_0}{r}=0,\ee
that is
\be \dot{A}_0=-\frac{32\kappa^{4}_{(3)}}{\kappa^{4}_{(4)}}r.\ee
This equation can easily be integrated to give
\be\label{sol-BTZ} A_0(r)=c-\frac{r^{2}}{2r_0^2},\ee
where $r_0^2=\frac{\kappa^{4}_{(4)}}{32\kappa^{4}_{(3)}}$ is the `crossover scale' and $c$ is an integration constant. This gives us an {\it exact}
three-dimensional black hole solution living on the 2-brane that can be expressed in the
usual form
\be
ds_{(3)}^{2}=-\left(c-\frac{r^{2}}{2r_0^2}\right)dt^{2}+
\left(c-\frac{r^{2}}{2r_0^2}\right)^{-1}dr^2+r^2d\theta^2.
\ee
{For $c=1$ we have just a de Sitter spacetime, whereas for $c=1-8G_{3}M$ we have a three-dimensional de Sitter - Schwarzschild ($SdS_3$) black hole, a conical singularity with associated mass $M$ \cite{mann}. We also define $\kappa^2_{(3)}\equiv 8\pi G_{3}$.  Assuming that our metric (\ref{4d-metric}) now describes a four-dimensional flat space with two timelike coordinates $t$ and $z$ which provides an $AdS_3$ embedding results  in the Eq. (\ref{eq.4d}) with a changed sign 
\be \label{eq.4d-II}
\frac{1}{2}\left(\frac{A'}{A}\right)^{2}-\frac{1}{r}\dot{A}=0.\ee
As a consequence our previous solution changes and now reads
\be
ds_{(3)}^{2}=-\left(c+\frac{r^{2}}{2r_0^2}\right)dt^{2}+
\left(c+\frac{r^{2}}{2r_0^2}\right)^{-1}dr^2+r^2d\theta^2.
\ee
For $c=-8G_{3}M$ this describes BTZ black hole solution embedded into the 2-brane. As expected the geometry asymptotically describes an $AdS_3$ space.
This metric is of the same type of BTZ black holes first treated in the Refs.~\cite{mann,banados, btz} with an important exception. }
The novelty here is that the black hole solution does not require a cosmological constant {\it a priori}. Instead, the induced curvature on the 2-brane
gives rise to a `cosmic fluid' that plays the role of a `cosmological constant'. Notice that for $\kappa^{2}_{(4)}\gg \kappa^{2}_{(3)}$ there is a four-dimensional gravity dominance over the three-dimensional gravity, such that 
$r_0$ is very large and then the solution (\ref{sol-BTZ}) approaches the solution of a flat space. However in four-dimensional gravity this is not the only vacuum solution. Rather, there is also the Schwarzschild solution, whose asymptotic behavior goes like $1/r$. 
We shall turn to this point shortly in the realm of the NMG \cite{bht}.

\section{Black Hole Solutions with Induced Scalar Massive Gravity on a 2-brane in 4D Flat Spacetime}

Let us now address the issue of having induced three-dimensional NMG on the 2-brane described by the following action \cite{bht}
\be
S_{(3)}=\frac{1}{2\kappa_{(3)}^{2}}\int
d^{3}x\sqrt{|q|}\left(R^{(3)}+\frac{1}{m^{2}}K\right),\ee
where $m^2$ is the mass of the three-dimensional gravity and
\be K=R^{(3)}_{\mu\nu}R^{(3)\mu\nu}-\frac{3}{8}R^{{(3)}\,2}.\ee
On the 2-brane we can write
\be G_{ab}+\frac{1}{2m^2}K_{ab}=\kappa_{(3)}^{2}U_{ab},\ee
where $U_{a}^{b}=\delta(z)\ {\rm diag\ }(-\rho_{curv},p_{curv},p_{curv},0)$. The tensor $K_{ab}$ is defined in terms of Levi-Civita covariant derivative, $R$ and $R_{ab}$ --- see Ref.~\cite{bht} for explicit form. Now using again the 
four-dimensional spacetime metric (\ref{4d-metric}) to have the induced three-dimensional metric at $z=0$ we find 
\ben
&&\rho_{curv}=-U_{t}^{t}=\frac{1}{\kappa^{2}_{(3)}}\Big[\frac{1}{2r}\frac{dA_0}{dr}-\frac{1}{8m^2}\frac{dA_0}{dr}\frac{d^3A_0}{dr^3}-
\frac{1}{4m^2}A_0\frac{d^4A_0}{dr^4}\nonumber\\
&-&\frac{1}{4m^2}\frac{A_0}{r}\frac{d^3A_0}{dr^3}-\frac{1}{8m^2}\frac{1}{r}\frac{dA_0}{dr}\frac{d^2A_0}{dr^2}
+\frac{1}{16m^2}\left(\frac{d^2A_0}{dr^2}\right)^2\Big],
\een
and by using the relation (\ref{Kcurv}), i.e.,
\be\frac{A'(0^+)}{A_0}=-\frac{\kappa^{2}_{(4)}}{2}\rho_{curv},\ee
we find the following important equation
\ben
&&\frac{A'(0^+)}{A_0}=-\frac{\kappa^{2}_{(4)}}{4\kappa^{2}_{(3)}}\Big[\frac{1}{r}\frac{dA_0}{dr}-\frac{1}{4m^2}\frac{dA_0}{dr}\frac{d^3A_0}{dr^3}-
\frac{1}{2m^2}A_0\frac{d^4A_0}{dr^4}\nonumber\\
&-&\frac{1}{2m^2}\frac{A_0}{r}\frac{d^3A_0}{dr^3}-\frac{1}{4m^2}\frac{1}{r}\frac{dA_0}{dr}\frac{d^2A_0}{dr^2}
+\frac{1}{8m^2}\left(\frac{d^2A_0}{dr^2}\right)^2\Big].\;\;\
\een
We are now able to use the equation (\ref{eq.4d}) with the boundary condition at $z=0$ 
\be\label{eq-New}
\frac{1}{2}\left(\frac{A'(0^+)}{A_0}\right)^2+\frac{1}{r}\frac{dA_0}{dr}=0,
\ee
such that
\ben\label{eqAm}
&&r_0^2\Big[\frac{1}{r}\frac{dA_0}{dr}-\frac{1}{2m^2}\left(\frac{1}{2}\frac{dA_0}{dr}+
\frac{A_0}{r}\right)\frac{d^3A_0}{dr^3}-\frac{1}{2m^2}A_0\frac{d^4A_0}{dr^4}\nonumber\\
&-&\frac{1}{4m^2}\frac{1}{r}\frac{dA_0}{dr}\frac{d^2A_0}{dr^2}
+\frac{1}{8m^2}\left(\frac{d^2A_0}{dr^2}\right)^2\Big]^2
+\frac{1}{r}\frac{dA_0}{dr}=0,
\een
where $r_0^2=\frac{\kappa^{4}_{(4)}}{32\kappa^{4}_{(3)}}$ as in the previous case.

Let us try to find a solution to this equation assuming the following Ansatz
\be\label{sol-type}
A_0(r)=\frac{a}{r^n}+br^2+c.
\ee
{We have found a class of  {\it exact} non-trivial solutions for $n=0$, assuming $r_{0-}^2 = 8/\alpha^2(\alpha^2+4)^2m^2$ and  $r_{0+}^2 = 8/(\alpha^2+8)(\alpha^2+4)^2m^2$, by properly using Eq.(\ref{eq.4d-II}) in place of Eq.(\ref{eq-New}), being $c$ a constant, such that our solutions become
\be\label{alpha-BTZ}
A_0^{(\alpha)}(r)=c\pm \frac{8}{(4+\alpha^2)^2}\frac{r^2}{r_{0\pm}^2}.
\ee
Recall that the case with plus sign and $c<0$ corresponds to BTZ solutions \cite{mann}. We remark that the above result suggests the existence of independent $\alpha$-type three-dimensional black holes with non-coincident horizons $r_\alpha=(4+\alpha^2)r_{0\pm}/2\sqrt{2}$, where $\alpha>0$.}

On the other hand, one can also explore an approximate solution of the type (\ref{sol-type})  with $n\neq0$ for a {\it near horizon} regime, say, around the largest horizon $r\simeq r_\alpha\simeq r_0$ and large $r_0$. Thus, we can write 
\be
A_0^{(\alpha)}(r)\simeq\frac{a}{r_\alpha^n}+c,
\ee
that approaches the constant $c$ for $a/r_\alpha^n\ll1$. For a finite value of $a$ this is particularly true for massless 
gravitons, $m\to0$, because $r\simeq r_\alpha\simeq r_0\sim 1/m\to\infty$. In the usual four-dimensional theory of gravity
the graviton is massless and the Newtonian potential goes like $1/r$. This allows us to fix $n=1$, $a=-GM$ and $c=1$ to find the Schwarzschild regime for very large three-dimensional black hole horizons $r_\alpha\simeq r_0$
\be
A_0^{(2)}(r)\simeq 1-\frac{GM}{r},
\ee
with $r\simeq r_0$; recall that $r_0^2={\kappa^{4}_{(4)}}/{32\kappa^{4}_{(3)}}$. Thus, this also signalizes that in this regime the four-dimensional gravity coupling
$\kappa^2_{(4)}\equiv 8\pi G$ is dominating over the three-dimensional gravity part controlled by the coupling $\kappa_{(3)}$ in our Lagrangian. This, of course, happens because we are probing distances so large as the crossover scale, so it is not surprising that the three-dimensional solutions approaches a four-dimensional solution. However, it is not necessary neither ask for a $r$ varying in the same way as $r_0$ nor assume $r_0$ to be very large. Indeed, as we show below, BTZ or three-dimensional de Sitter - Schwarzschild ($SdS_3$) solutions can flow to higher dimensional solutions as we probe sufficient large distances along the 2-brane.

Since we are working with induced gravity on a 2-brane embedded into a four-dimensional spacetime it is interesting to know whether two-dimensional gravitational solutions have higher dimensional (i.e. four-dimensional) deviations from the bulk at some scale. In other words, we shall see how BTZ/$SdS_3$ black hole solutions can receive deviations such that at some sufficient large distance they look four-dimensional Schwarzschild black holes --- a similar investigation was carried out in \cite{ehm}. Thus,  we now address the issue of finding local black holes solutions that can flow to each other as one goes from small ($r\ll r_0$)  through large ($r\gg r_0$) distances in the realm of NMG induced on a 2-brane embedded in a flat four-dimensional spacetime. 

\begin{figure}
\centerline{\includegraphics[width=7cm]{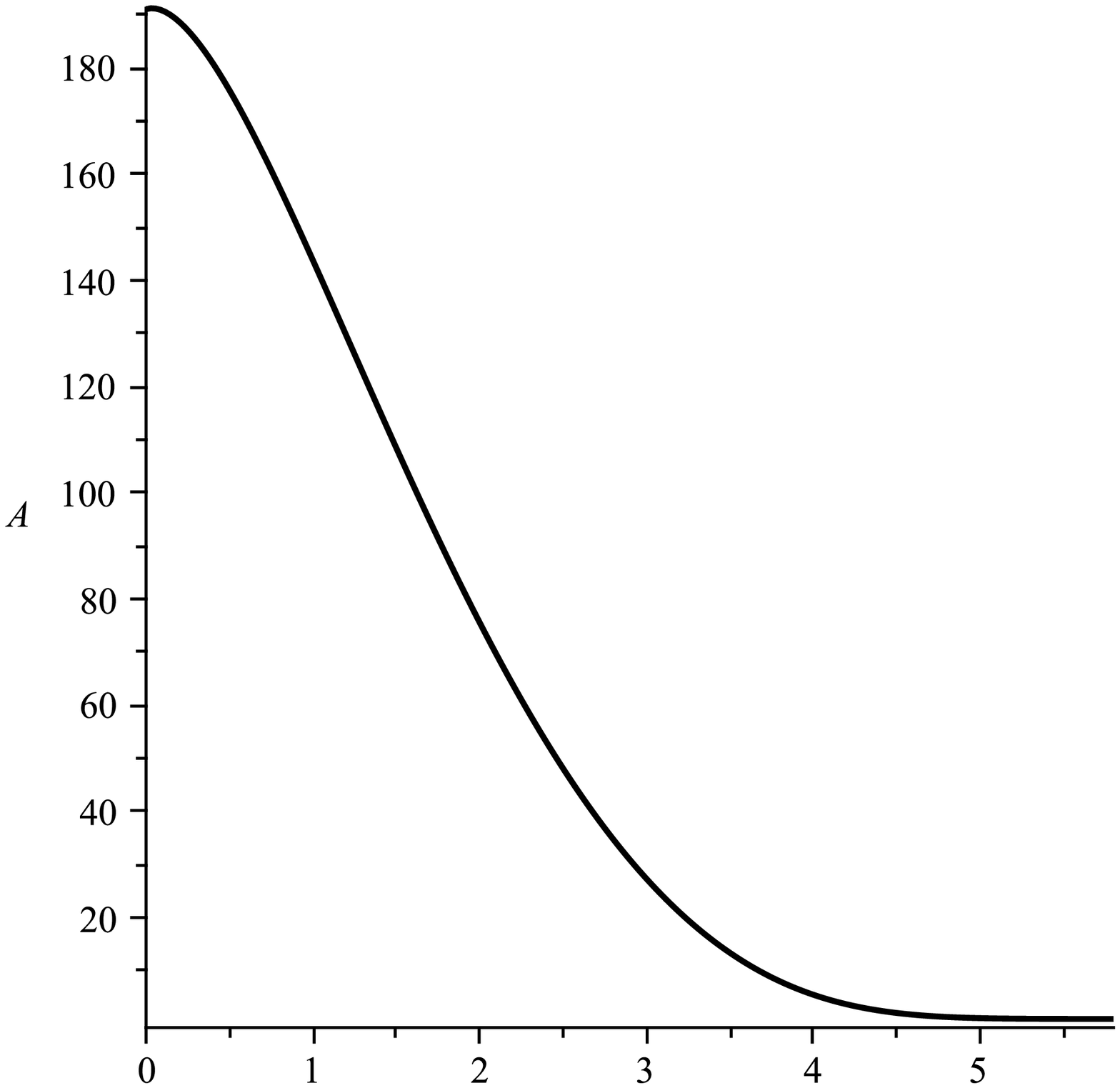}
\includegraphics[width=7cm]{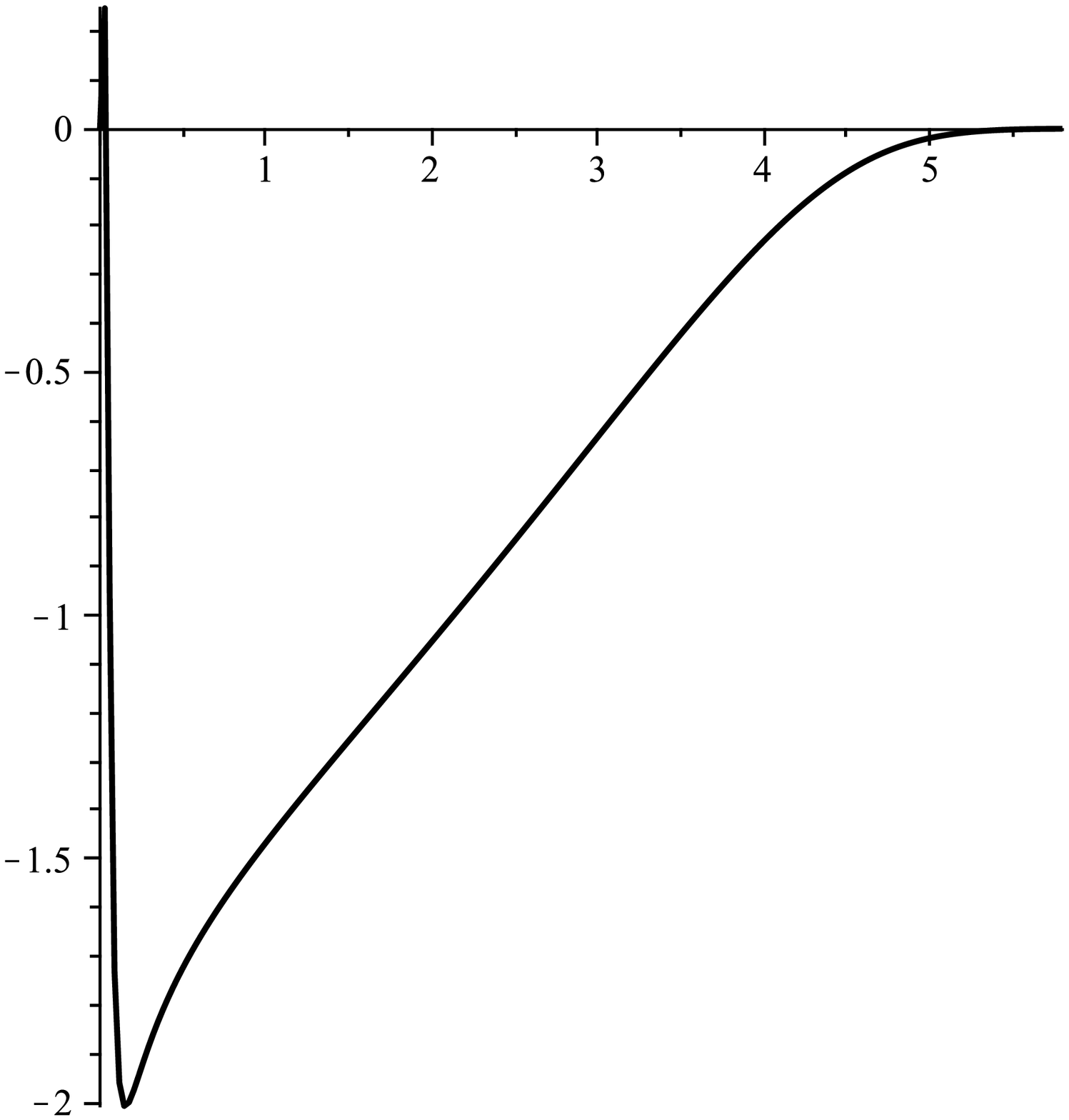}}
\caption{Left panel: The warp factor $A(\tilde{r})$ for small distances. We find a local $SdS_3$ for $\tilde{r}\simeq0$.  Right panel: The precise power $d(\tilde{r}\simeq0)\simeq-2$ of the `potential' $A(\tilde{r})-c\sim-1/\tilde{r}^{-2}$, being $c=191.59$.}
\label{fig1}
\end{figure}

\begin{figure}
\centerline{\includegraphics[width=7cm]{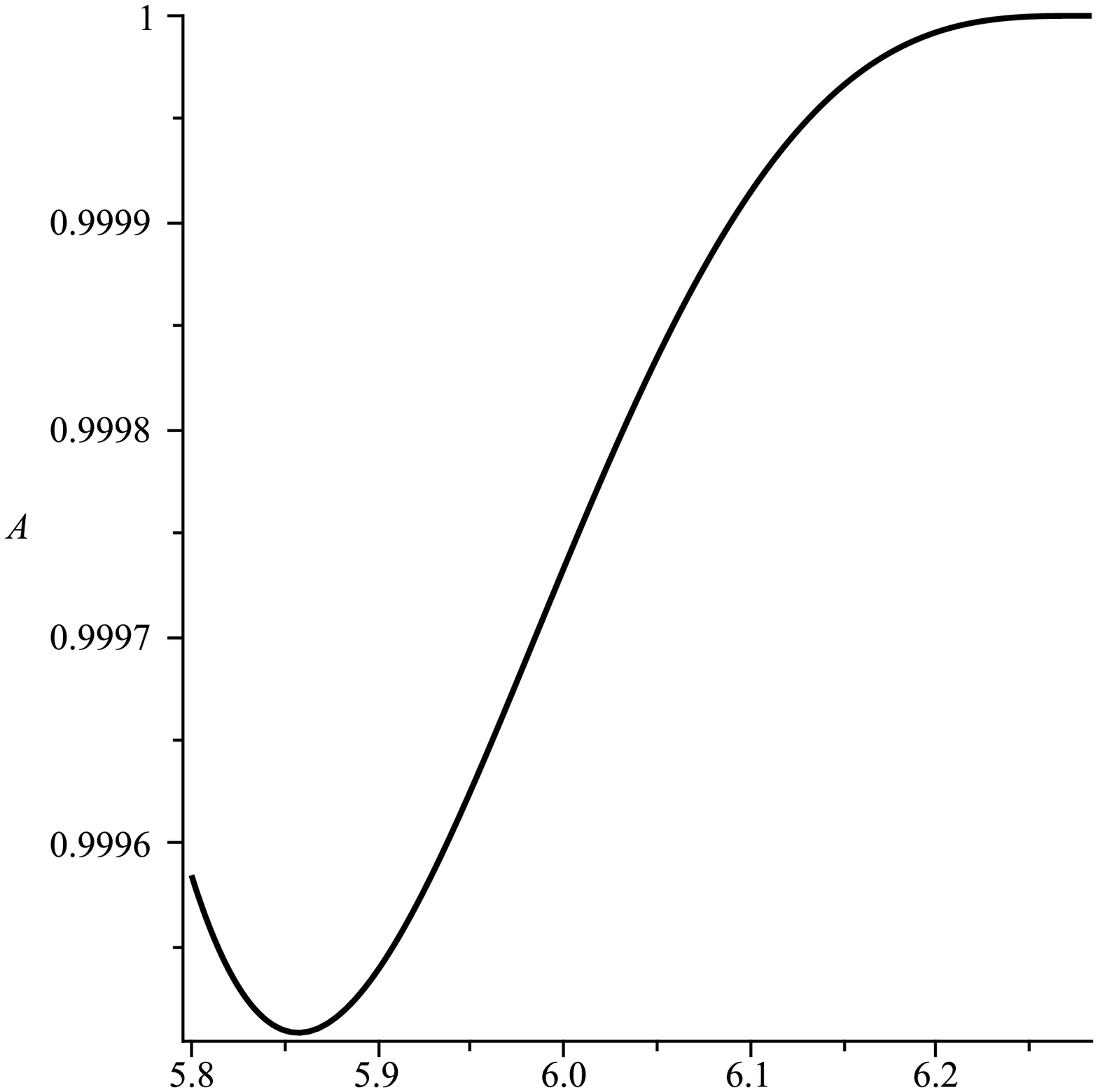}
\includegraphics[width=7cm]{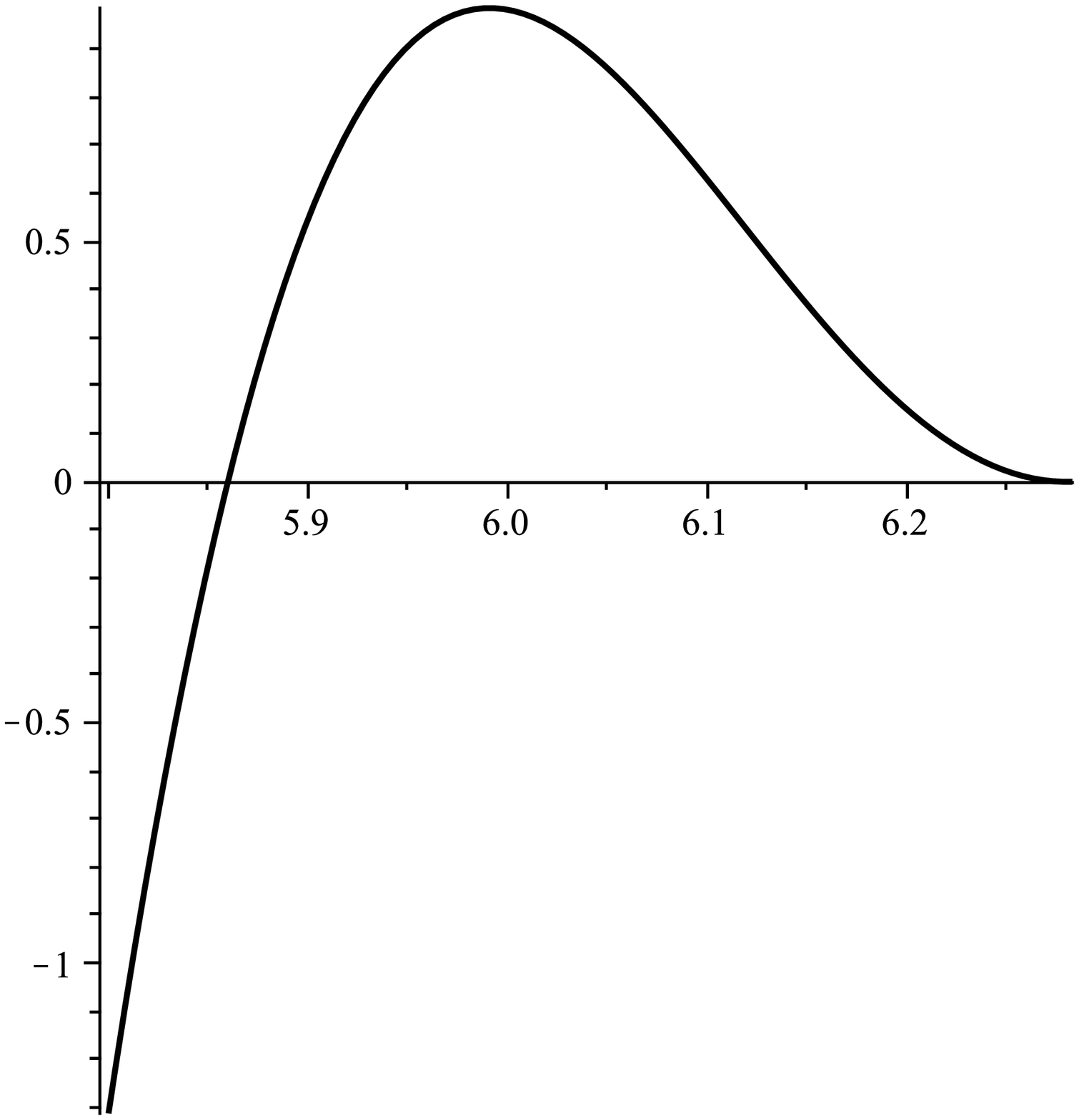}}
\caption{Left panel: The warp factor $A(\tilde{r})$ for large distances. We approach a local Schwarzschild-like  solution for $\tilde{r}\simeq6$.  Right panel: The precise power $d(\tilde{r}\simeq6)\simeq1$ of the `potential' $A(\tilde{r})-c\sim-1/\tilde{r}$, being $c=1.013$.}
\label{fig2}
\end{figure}

\begin{figure}
\centerline{
\includegraphics[width=7cm]{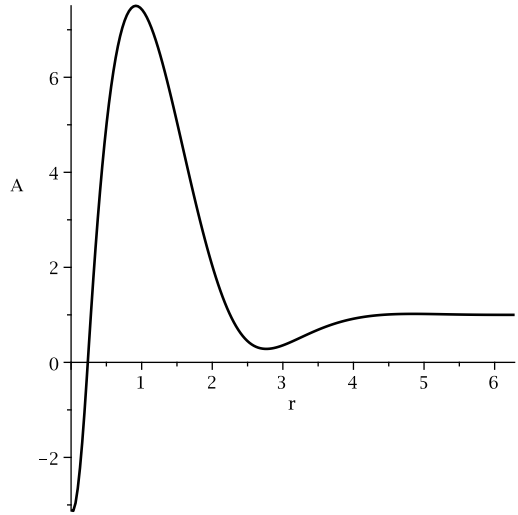}}
\caption{The warp factor $A(\tilde{r})$ for the transition BTZ - $SdS_3$ and $SdS_3$-Schwarzschild-like regimes.}
\label{fig5}
\end{figure}

For this, we content ourselves with numerical solutions of the differential equation (\ref{eqAm}). We work with the dimensionless variables $\tilde{r}=r/r_0$ and $\alpha=m r_0$, being $m$ and $r_0$ the mass and the crossover scale. The boundary conditions are such that we can pick a branch distinct of the one that gave us the exact BTZ/$SdS_3$ solutions (\ref{alpha-BTZ}). Notice that for BTZ/$SdS_3$ solutions the third and fourth derivatives are zero everywhere. So we have chosen the following boundary conditions: $A(\tilde{r}_\infty)=1, A'(\tilde{r}_\infty)=0, A''(\tilde{r}_\infty)=0, A'''(\tilde{r}_\infty)=1/10 $. The last two boundary conditions prevent BTZ/$SdS_3$ solutions at the scale $\tilde{r}_\infty$. The boundary $\tilde{r}_\infty$ is assumed to be a sufficiently large scale --- indeed, this is not necessary since several tests revealed that the position of this boundary is not relevant.  In our results we are assuming $0\leq\tilde{r}<\tilde{r}_\infty$ with $\tilde{r}_\infty=2\pi$ and $\alpha=6.7$. The choice of $\alpha$ just reflects on the position of the local BTZ/$SdS_3$ black hole. The results are shown in Figs.~\ref{fig1} and \ref{fig2} for $SdS_3$ case. The right panel of these figures show how the power $d(\tilde{r})$ of the `potential' $V(\tilde{r})\sim A(\tilde{r})-c$ changes at each point. Thus we define the useful quantity $d(\tilde{r})$ as the `effective dimension'
\begin{equation}
d(\tilde{r})=-\frac{d\ln{V(\tilde{r})}}{d\ln{\tilde{r}}},
\end{equation}
which is pretty similar to the definition of spectral dimension \cite{h1,h2,amb1,amb2,correia,Brito:2010by}. The Fig.~\ref{fig5} shows transition BTZ - $SdS_3$ and $SdS_3$-Schwarzschild-like regimes for $A'''(\tilde{r}_\infty)=-1/10$, $\alpha=25$ and $r_0=6$.

\section{Conclusions}
\label{conclu}

In this paper we have found three-dimensional black hole solutions in realm of pure and new massive gravity (NMG) induced on a 2-brane embedded in a flat four-dimensional spacetime. There is
no cosmological constant  neither on the brane nor on the four-dimensional bulk. Only gravitational fields are turned on and we indeed find vacuum solutions as black holes in 2+1 dimensions even in
the absence of any cosmological solution. In the first part of our investigation the induced gravity on the 2-brane without new massive gravity term (pure gravity) reproduces a BTZ/$SdS_3$ type black hole whose horizon has dependence 
with the ratio between the four- and three-dimensional gravitational couplings $\kappa_{(4)}$ and $\kappa_{(3)}$. This allows us to identify one crossover scale that controls how far  the
three- or four-dimensional gravity manifests on the 2-brane. In the second part we include the new massive gravity term. The mass of the gravitons now control the crossover scale. In this realm we still find exact
black hole solutions with distinct horizons. An effect can be 
identified as the probed distances on the 2-brane are around the crossover scale but with the crossover scale very large (this happens as the graviton mass approaches zero). It follows that in this regime the
three-dimensional black hole solution, as expected,  can approach a Schwarzschild black hole solution in 3+1 dimensions. Furthermore, even if the crossover scale is not necessary large, the numerical solutions indicate a more interesting effect.  Now, the local BTZ/$SdS_3$ solutions can flow to four-dimensional Schwarzschild like black holes, as one moves from small to large distances, which is clearly a higher dimensional manifestation on the 2-brane. This is similar to the effect of manifestation of extra dimensions for large probed distances
along the brane in the DGP scenario.

\acknowledgments

We would like to thank CNPq, CAPES, PNPD/PROCAD - CAPES for partial financial support.

\end{document}